\documentclass[fleqn,usenatbib]{mnras}

\usepackage{newtxtext,newtxmath}
\usepackage[T1]{fontenc}

\DeclareRobustCommand{\VAN}[3]{#2}
\let\VANthebibliography\thebibliography
\def\thebibliography{\DeclareRobustCommand{\VAN}[3]{##3}\VANthebibliography}

\usepackage{graphicx}	
\usepackage{amsmath}	

\title[3D O-shell burning in a rotating massive star]{A three-dimensional hydrodynamics simulation of oxygen-shell burning in the final evolution of a fast-rotating massive star
}

\author[T. Yoshida et al.]{
Takashi Yoshida,$^{1,2}$\thanks{E-mail: yoshida@yukawa.kyoto-u.ac.jp}
Tomoya Takiwaki,$^{3}$
David R. Aguilera-Dena,$^{4,5,6}$
Kei Kotake,$^{7}$
Koh Takahashi,$^{8}$ \newauthor
Ko Nakamura,$^{7}$
Hideyuki Umeda,$^{1}$
Norbert Langer,$^{5,6}$
\\
$^{1}$Department of Astronomy, Graduate School of Science, The University of Tokyo, 7-3-1 Hongo, Bunkyo-ku, Tokyo 113-0033, Japan\\
$^{2}$Yukawa Institute for Theoretical Physics, Kyoto University, Kitashirakawa Oiwakecho, Sakyo-ku, Kyoto 606-8502, Japan\\
$^{3}$Division of Science, National Astronomical Observatory of Japan, National Institutes for Natural Science, 2-21-1 Osawa, Mitaka, Tokyo 181-8588, Japan\\
$^{4}$Institute of Astrophysics, FORTH, Department of Physics, University of Crete, Voutes, University Campus, GR-71003 Heraklion, Greece\\
$^{5}$Argelander-Institut f\"ur Astronomie, Universit\"at Bonn, Auf dem H\"ugel 71, D-53121 Bonn, Germany\\
$^{6}$Max-Planck-Institute f\"ur Radioastronomie, Auf dem H\"ugel 69, D-53121 Bonn, Germany\\
$^{7}$Department of Applied Physics \& Research Institute of Stellar Explosive Phenomena, Fukuoka University, Fukuoka 814-0180, Japan\\
$^{8}$Max-Planck-Institute for Gravitational Physics, D-14476 Potsdam, Germany\\
}

\date{Accepted XXX. Received YYY; in original form ZZZ}

\pubyear{2021}

\begin{document}
\label{firstpage}
\pagerange{\pageref{firstpage}--\pageref{lastpage}}
\maketitle

\begin{abstract}
We perform for the first time a 3D hydrodynamics simulation of the evolution of the last minutes pre-collapse of the oxygen shell of a fast-rotating massive star.
This star has an initial mass of 38 M$_\odot$, a metallicity of $\sim$1/50 Z$_\odot$, an initial rotational velocity of 600 km s$^{-1}$, and experiences chemically homogeneous evolution.
It has a silicon- and oxygen-rich (Si/O) convective layer at (4.7--17)$\times 10^{8}$ cm, where oxygen-shell burning takes place.
The power spectrum analysis of the turbulent velocity indicates the dominance of the large-scale mode ($\ell \sim 3$), which has also been seen in non-rotating stars that have a wide Si/O layer.
Spiral arm structures of density and silicon-enriched material produced by oxygen-shell burning appear in the equatorial plane of the Si/O shell.
Non-axisymmetric, large-scale ($m \le 3$) modes are dominant in these structures.
The spiral arm structures have not been identified in previous non-rotating 3D pre-supernova models.
Governed by such a convection pattern, the angle-averaged specific angular momentum becomes constant in the Si/O convective layer, which is not considered in spherically symmetrical stellar evolution models.
Such spiral arms and constant specific angular momentum might affect the ensuing explosion or implosion of the star.

\end{abstract}




\section{Introduction}

Multidimensional hydrodynamics simulations of  convective motion in the last several minutes before the core collapse of massive stars have progressed significantly in recent years \citep[e.g.,][]{bazan98,meakin06,Couch15,bernhard16_prog,bernhard18_prog,Yoshida19,Yoshida21,Fields20,McNeill20,Yadav20}.
It has been pointed out that aspherical characteristics induced by the convective motion in the silicon (Si) or silicon- and oxygen-rich (Si/O) layer give a favorable condition for a powerful supernova (SN) explosion \citep{bernhard17,Bollig20}.

On the other hand, 1D evolution of rotating massive stars have been systematically studied \citep[e.g.,][]{Brott11,Ekstroem12,Limongi18} and SN progenitors evolved from rotating massive stars have been constructed \citep[e.g.,][]{Heger00,Hirschi04,Heger05,Limongi18}.
The characteristic evolution of fast-rotating stars such as chemically homogeneous evolution has been revealed \citep{Maeder87}.
The characteristics of the SN progenitor structure evolved from rotating stars and binary stars are drawing attention from the viewpoint of the progenitors of long gamma-ray bursts (GRBs) and superluminous (SL) SNe \citep[e.g.,][]{Yoon06,Woosley07,Woosley12}.

Concerning multidimensional simulations of the convective motion in rotating massive stars, the effects of rotation on convective carbon, oxygen, and silicon-shell burning during the late stage of evolution in a 20 M$_\odot$ star have been investigated using 2D hydrodynamics simulations \citep{Chatzopoulos16}.
They found that adding modest amounts of rotation has little impact on the character of the convection.
However, in order to find non-axisymmetrical rotation effects in convective layers, it is necessary to perform 3D hydrodynamics simulations.

Recently, the evolution of low-metallicity fast-rotating massive stars, which are proposed as progenitors of SL SNe and long GRBs, have been systematically studied \citep{Aguilera-Dena18,Aguilera-Dena20}.
Some of these stars have a wide Si/O convective layer similar to the convective layer studied in \citet{bernhard16_prog,Yadav20,Yoshida19}, and \citet{Yoshida21}.
Thus, we perform for the first time, a 3D hydrodynamics simulation of the Si/O convective layer in a fast-rotating massive star during the last $\sim$90 s before its core collapse.
We show the effects of rotation on the distributions of the Si mass fraction, density, and the rotational kinetic energy density.
We also show the power spectra of degree $\ell$ for the radial turbulent velocity induced by the turbulence and that of order $m$ for the density in the {\it x--y} plane induced by rotation.
We further discuss the effects of convection on the radial profiles of angle-averaged angular velocity and specific angular momentum.

\section{Model}
\label{sec:star_model} 

We adopt a pre-collapse fast-rotating massive star model of \citet{Aguilera-Dena20} as initial configuration for the 3D hydrodynamics simulation.
The initial mass, the metallicity, and the initial rotating velocity of the star are 38 M$_\odot$, $Z$=0.0034 ($\sim$1/50 Z$_\odot$), and 600 km s$^{-1}$, respectively.
The evolution was calculated with the assumptions of the B series in \citet{Aguilera-Dena18}, i.e., the Tayler--Spruit dynamo is employed to calculate the transport of angular momentum and the diffusion coefficient due to rotational mixing is enhanced by a factor of 10 \citep[cf.][]{Hastings20} using the Modules for Experiments in Stellar Astrophysics code -- in its version 10398 \citep{Paxton11,Paxton13,Paxton15,Paxton18}.
This star evolves quasi-chemically homogeneously in the hydrogen and helium burning.
All of the hydrogen and almost all of helium in the envelope have been lost during the evolution.
The final mass is 28.68 M$_\odot$.

We note that the initial rotating velocity of 600 km s$^{-1}$ corresponds to the fastest rotating star in 30 Dor in the Large Magellanic Cloud \citep{Ramirez-Agudelo13}.
Rotating massive stars experiencing chemically homogeneous evolution are a candidate of SL SNe and long GRBs.
The event rates of Type Ic SL SNe and long GRBs are $\sim$10$^{-4}$ and $10^{-3}$ of the rate of core-collapse SNe \citep[e.g.,][]{Moriya18}.
Although the population of fast-rotating stars would be very small, these stars are important as candidates of Type Ic SL SNe and long GRBs.

\begin{figure}
\begin{center}
\includegraphics[width=.9\columnwidth]{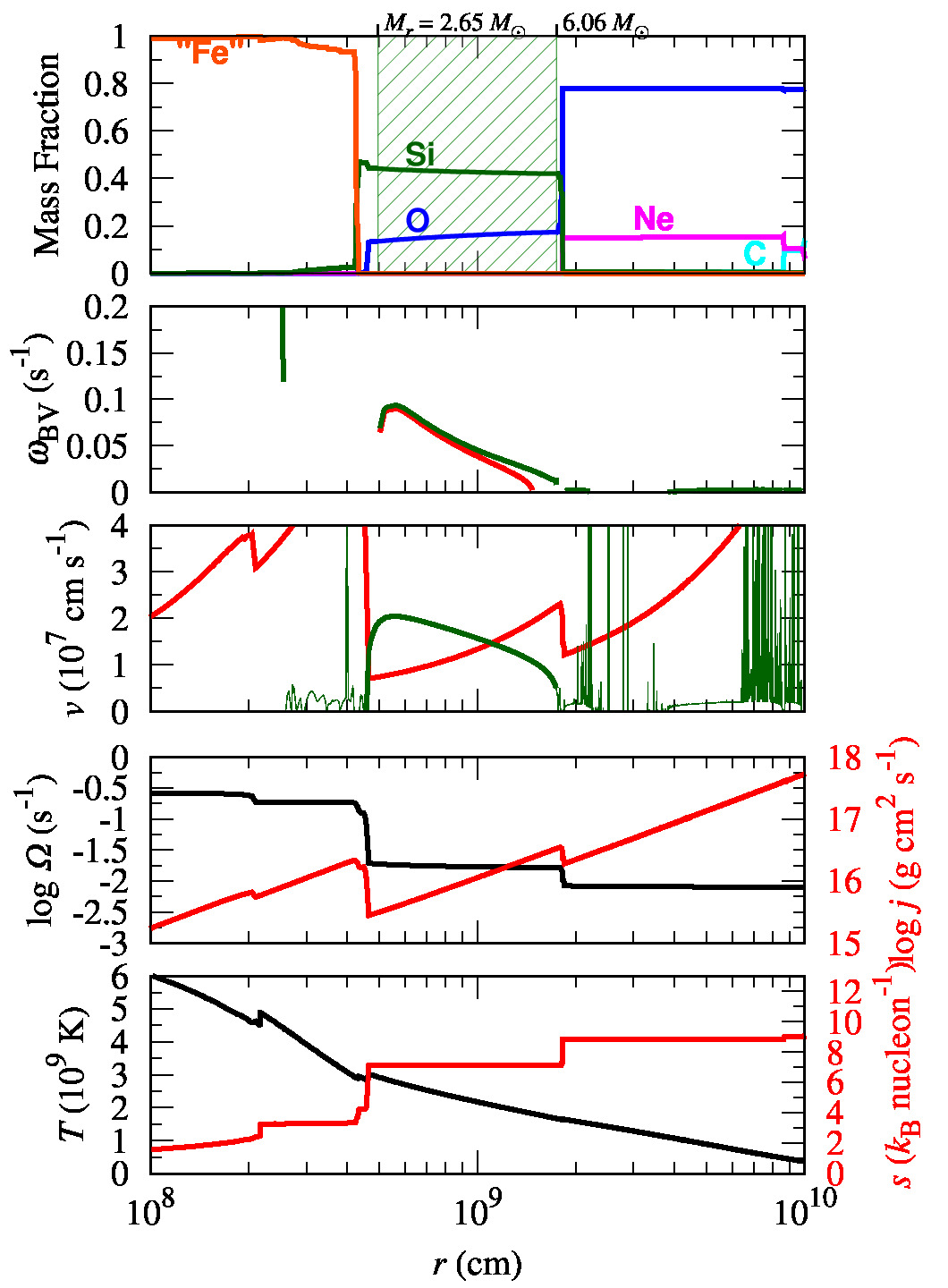}
\end{center}
\caption{Top panel: Abundance profiles at the initial time of the 3D hydrodynamics simulation.
Second panel: The profile of the BV frequency $\sqrt{-N^2}$ (green curve) taken from the 1D simulation.
The red curve takes into account the rotation effect $\sqrt{-N_{\rm rot}^2}$.
Third panel: The profiles of convective velocity deduced using mixing length theory (green curve) and the polar-angle-averaged rotation velocity (red curve) taken from the 1D simulation.
Fourth panel: The profiles of angular velocity (black curve) and specific angular momentum (red curve).
Bottom panel: The profiles of temperature (black curve) and entropy per baryon number (red curve), where $k_{\rm B}$ is Boltzmann constant.
}
\label{fig:1dmodel}
\end{figure}

We use the stellar structure at $\sim$100 s before core collapse (defined as the point where the infall velocity in the core reaches 1000 km s$^{-1}$), where the central temperature and density are $\log T_{\rm C} = 9.827$ and $\log \rho_{\rm C} = 8.642$, respectively.
Fig. \ref{fig:1dmodel} shows the radial profiles of abundances, the Brunt--V\"ais\"al\"a (BV) frequency, angular velocity, specific angular momentum, temperature, and entropy.
The BV frequency $\omega_{\rm BV}$ is calculated as $\sqrt{-N^2}$ where
$N^2 = \frac{G M_r}{r^2}\frac{\delta}{H_P}(\nabla_{\rm ad} - \nabla + \frac{\varphi}{\delta} \nabla_{\mu}),$
where $G$ is the gravitational constant, $M_r$ is mass coordinate, $r$ is radius, $\delta = -(\partial \ln \rho/\partial \ln T)_{P,\mu}$, $P$ is pressure, $\mu$ is mean molecular weight, $H_P$ is pressure scale height, $\nabla_{\rm ad} \equiv (\partial \ln T/\partial \ln P)_{\rm ad}$ is the adiabatic temperature gradient, $\nabla = {\rm d} \ln T/{\rm d} \ln P$, $\varphi \equiv (\partial \ln \rho/\partial \ln \mu)_{P,T}$, and $\nabla_\mu = {\rm d}\ln \mu/{\rm d}\ln P$.
We also show the BV frequency including the rotation effect defined as $\sqrt{-N_{\rm rot}^2}$ where
$
    N_{\rm rot}^2 = N^2 + \frac{1}{\varpi^3}\frac{{\rm d}(\Omega^2\varpi^4)}{{\rm d} \varpi} \sin\theta,
$
where $\Omega$ is angular velocity, $\varpi = r \sin \theta$, $\theta$ is polar angle, and we assume $\sin \theta = 0.5$ in this figure.

This stellar model has an Si/O convective layer in the range (4.7--17) $\times 10^{8}$ cm (2.65--6.06 M$_\odot$).
The mass fractions of silicon and oxygen are $\sim$0.42--0.44 and 0.13--0.17, respectively, in this layer.
The BV frequency is $\sim$0.1 at its maximum.
When we consider the effect of rotation, the BV frequency slightly decreases (see the red curve).
This layer rotates with an almost constant angular velocity of $\Omega \sim 0.017$ s$^{-1}$.
The constant angular velocity profile in this layer has been established by the convection in the 1D calculation.
The temperature at the bottom of this layer is about $3 \times 10^{9}$ K and oxygen burns vigorously.
The entropy profile is almost constant in this layer due to convection.

Note that there is an oxygen- and neon-rich (O/Ne) layer above the Si/O layer.
This layer is also convectively unstable but the BV frequency is much smaller than in the Si/O layer.
In our simulation, we do not see any strong convective features in this layer.
So, this layer is not discussed further in this letter.

We perform a 3D hydrodynamics simulation employing the numerical schemes of \citet{Yoshida21}.
We use the 3DnSEV code, which solves Newtonian hydrodynamics equations.
We employ a spherical polar grid with a resolution of $n_r \times n_\theta \times n_\phi = 512 \times 64 \times 128$ zones.
The simulation extends from the center of the star to the outer boundary of $10^{10}$ cm.
Gravity is assumed as a 1D monopole, gravitational potential.
When we map the 1D structure of the rotating star model into the initial structure of the 3D simulation, we assume spherically symmetrical structure.
We set the angular velocity distribution as shellular rotation.

\section{Results}

The 3D simulation of the fast-rotating 38 M$_\odot$ model is performed for 91.6 s until
$\log T_{\rm C} = 10$.
Similar to the non-rotating models in the literature, large-scale turbulent flows develop outward from the bottom of the Si/O layer and reach the outer boundary after $\sim$50 s.
Then, oxygen-shell burning becomes more violent after $\sim$70 s due to the contraction of the burning layer.
Silicon-rich material is produced at the bottom and is carried outward by turbulent flows.
Owing to the shell burning, the turbulent Mach number increases.
When we stop the simulation, the maximum turbulent Mach number is $\langle Ma^2 \rangle^{1/2} = 0.135$ at $r = 5.7 \times 10^{8}$ cm (see Equation (2) in \citet{Yoshida19} for the definition).
The turnover time-scale of the convective motion is $\sim$18 s.
A large-scale mode ($\ell = 3$) of the radial turbulent velocity dominates in a wide range of the convective layer (see \S 3.2 for details).
These turbulence features are similar to those of models 25M in \citet{Yoshida19} and 22L in \citet{Yoshida21}.

Furthermore, we find rotation effects in the large-scale turbulent flows in the Si/O layer.
Hereafter, we show these effects in the profiles of the silicon mass fraction, the rotational kinetic energy, and the density.
We also show the radial profiles of the angle-averaged angular velocity and specific angular momentum at the last time step.

\subsection{Spiral arms}

\begin{figure*}
    \centering
    \includegraphics[width=2.0\columnwidth]{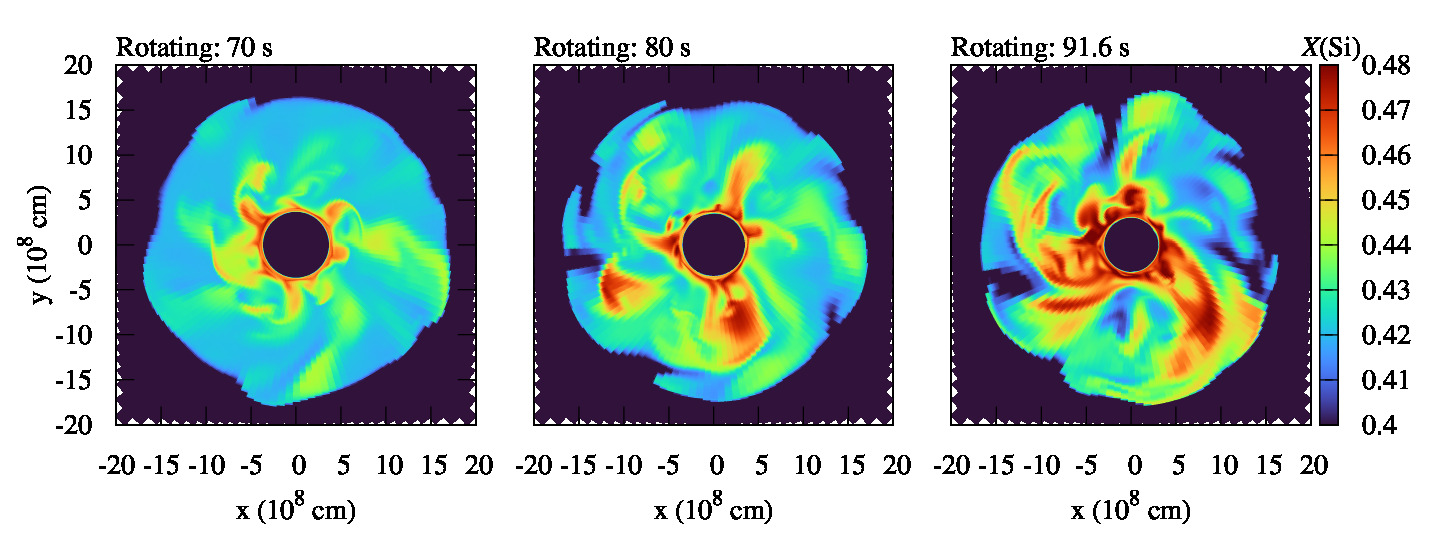}
\caption{Slices in the {\it x--y} plane showing the silicon mass fraction at 70, 80, and 91.6 s from left to right. 
}
\label{fig:2d_mfSi}
\end{figure*}

First, we show rotation effects in the evolution of the silicon mass fraction profile of the Si/O layer.
Fig. \ref{fig:2d_mfSi} shows the silicon mass fraction in the {\it x--y} plane.
We see that the silicon-rich material shown as reddish regions appears from the bottom of the Si/O layer at 70 s (left-hand panel).
The silicon-rich material is produced through oxygen-shell burning after $\sim$70 s.
These Si-rich regions grow outward and finally produce spiral-arm structures in the southeast and west directions at 91.6 s (right-hand panel).
The outward growth of the Si-rich region is due to the turbulent motion induced by the shell burning.
The spiral-arm structures are a rotation effect seen in a convective layer of our fast-rotating massive star.

\begin{figure}
    \centering
     \includegraphics[width=0.8\columnwidth]{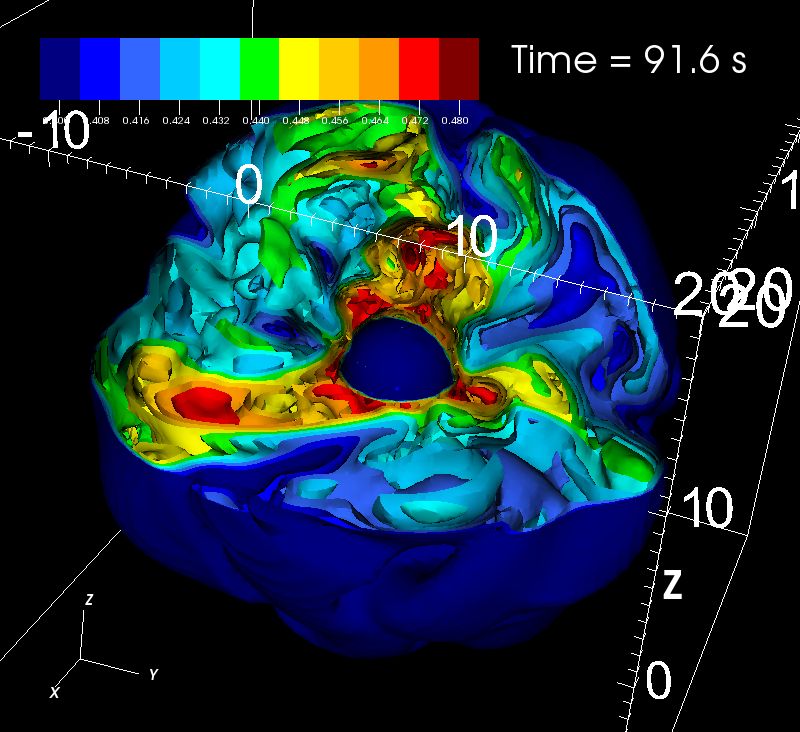}\\
\caption{
3D contours of the silicon mass fraction at 91.6 s.
}
\label{fig:3d_mfSi}
\end{figure}

We also see a spiral-arm structure in the 3D contours of Fig. \ref{fig:3d_mfSi}.
There is a Si-rich arm region in the equatorial plane.
This corresponds to the arm in the south-east region in the right-hand panel of the {\it x--y} plane.
On the other hand, we see another Si-rich region in the northern direction in the {\it x--z} plane.
This region also indicates an outflow of silicon-rich material.
This outflow looks like a blob rather than a spiral-arm.
It is a convective (turbulent) flow produced through the oxygen-shell burning and is not affected by the rotation.

We see the development of spiral-arms and turbulent blobs of silicon-rich material in the Si/O layer.
However, the time from the ignition of the oxygen-shell burning to the core collapse is too short to homogenize the abundance distribution throughout the Si/O layer.
Although the oxygen abundance is reduced in the Si/O layer, the region showing an angle-averaged oxygen mass fraction less than 0.13 (the lowest value in the Si/O layer at initial time of the simulation) is limited to $r < 9 \times 10^{8}$ cm.

\begin{figure}
    \centering
    \includegraphics[width=0.495\columnwidth]{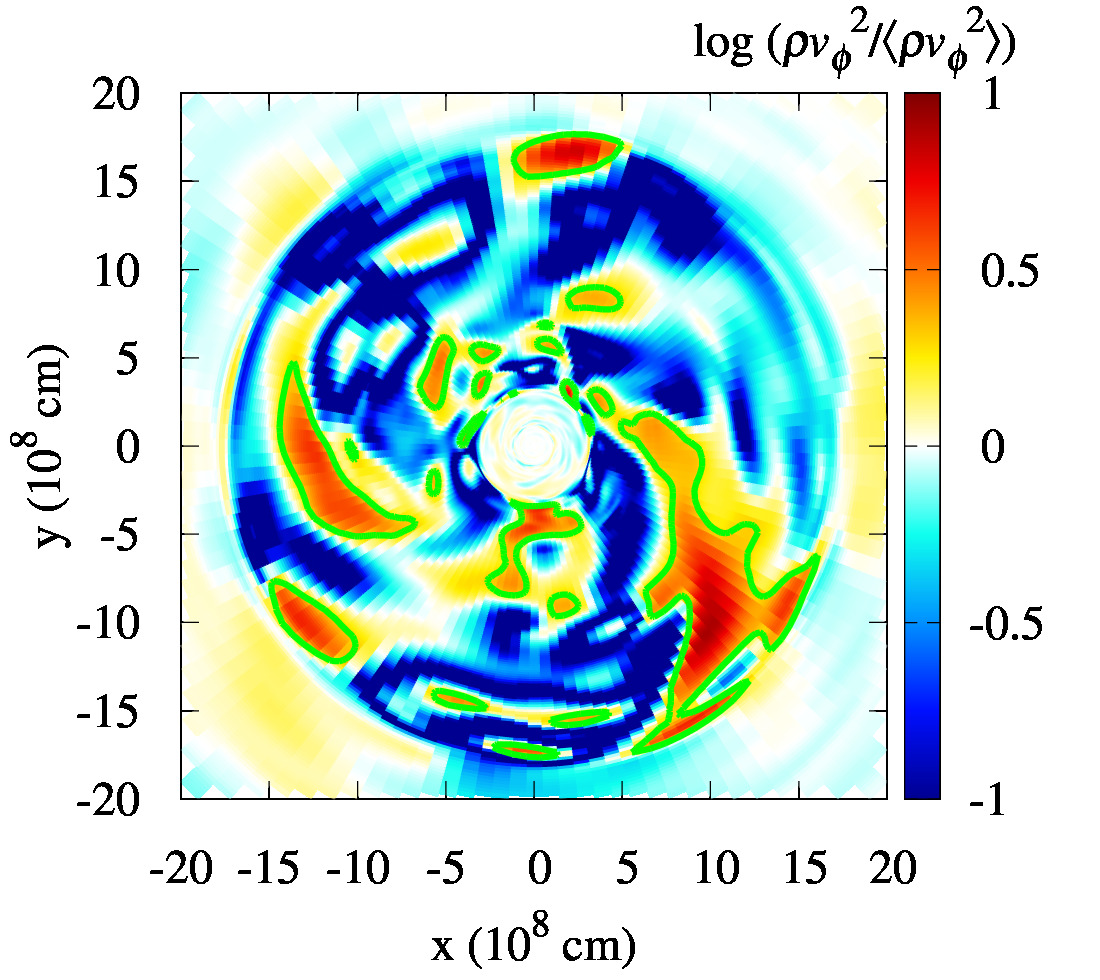}
    \includegraphics[width=0.495\columnwidth]{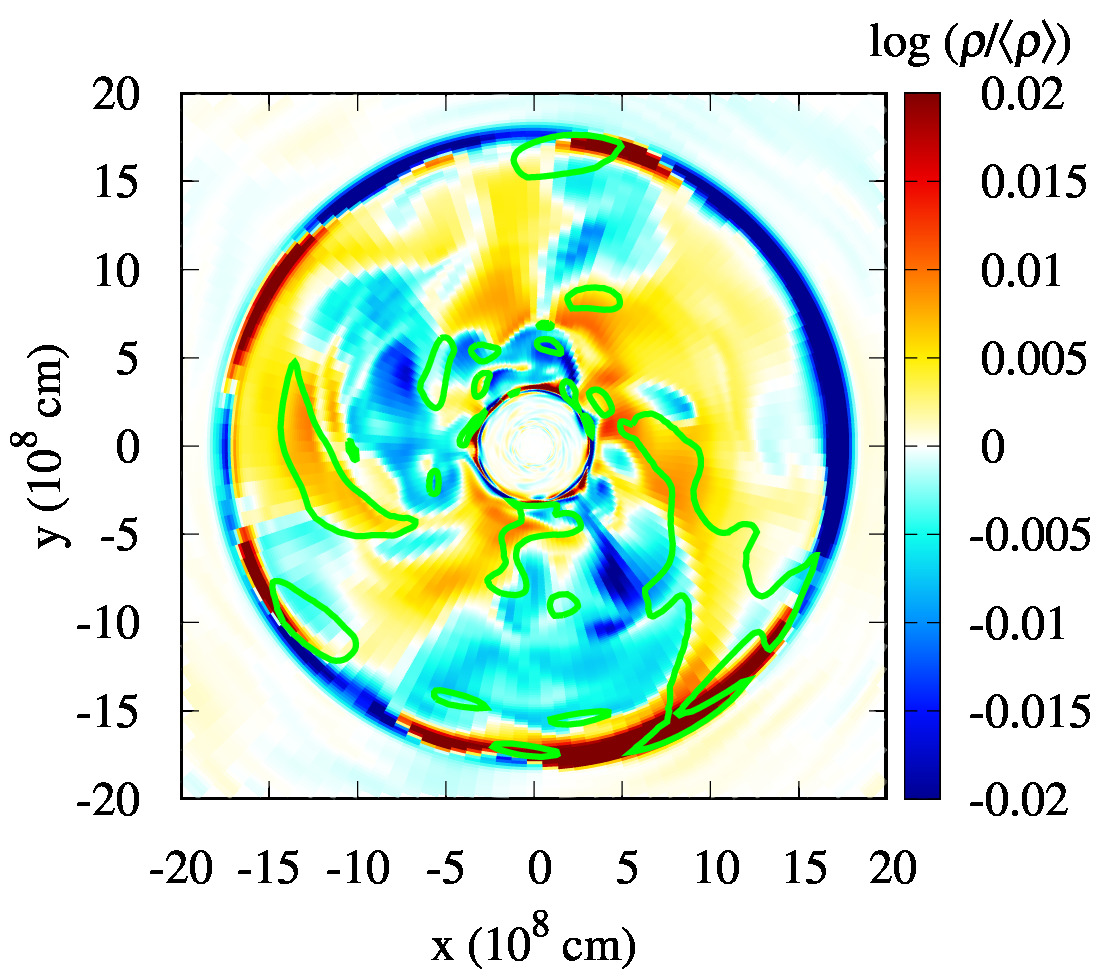}
\caption{
Slices in the {\it x--y} plane showing the fluctuations in the rotational kinetic energy density ($\log (\rho v_{\phi}^2/\langle \rho v_{\phi}^2 \rangle)$ (left-hand panel) and the density $\log (\rho/\langle \rho \rangle)$ (right-hand panel) at 91.6 s.
In the both panels, green curves indicate contours of $\log (\rho v_{\phi}^2/\langle \rho v_{\phi}^2 \rangle) = 0.3$.
}
\label{fig:2d_e3_rho}
\end{figure}

Next, we show rotation effects on turbulent motion in the Si/O layer.
To quantify the rotational effects, we focus on the profiles of the rotational kinetic energy density and the density.
When fluctuations in these quantities caused by rotation are present, we present their value along the azimuthal angle in the {\it x--y} plane.
A quantity averaged by azimuthal angle in the {\it x--y} plane $\langle q \rangle$ is calculated as
$
    \langle q \rangle = (1/2\pi) \int q d\phi.
$

Fig. \ref{fig:2d_e3_rho} shows slices in the {\it x--y} plane for the fluctuations of the rotational kinetic energy density $\log (\rho v_\phi^2 / \langle \rho v_\phi^2 \rangle )$ and density $\log (\rho /\langle \rho \rangle)$ at 91.6 s.
In the rotational kinetic energy density profile, there are reddish spiral arms in the southeast and west directions and blueish regions in the north and south directions (see the left-hand panel).
Such spiral-arm structures have not been identified in previous 3D SN progenitor models.

For the slice in the {\it x--y} plane of the density fluctuation, we see an enhancement in the east and west directions and a reduction in the north and south direction.
We also draw green contours of the enhancement of the rotational kinetic energy density on the density color map.
Although the locations of the enhancement in the density profile are not similar to those in the rotational energy density, we may see fluctuations of low-order mode such as $m \sim 2$ by considering the power spectrum analysis in the {\it x--y} plane as discussed below.

\subsection{Power spectrum}

In this subsection, we analyze the power spectrum of degree $\ell$ mode for the radial turbulent velocity to find the characteristics of turbulent motion in the Si/O layer, and that of order $m$ mode for the density fluctuations in the $x$-$y$ plane to find the effect of rotation.

\begin{figure}
    \centering
    \includegraphics[width=0.9 \columnwidth]{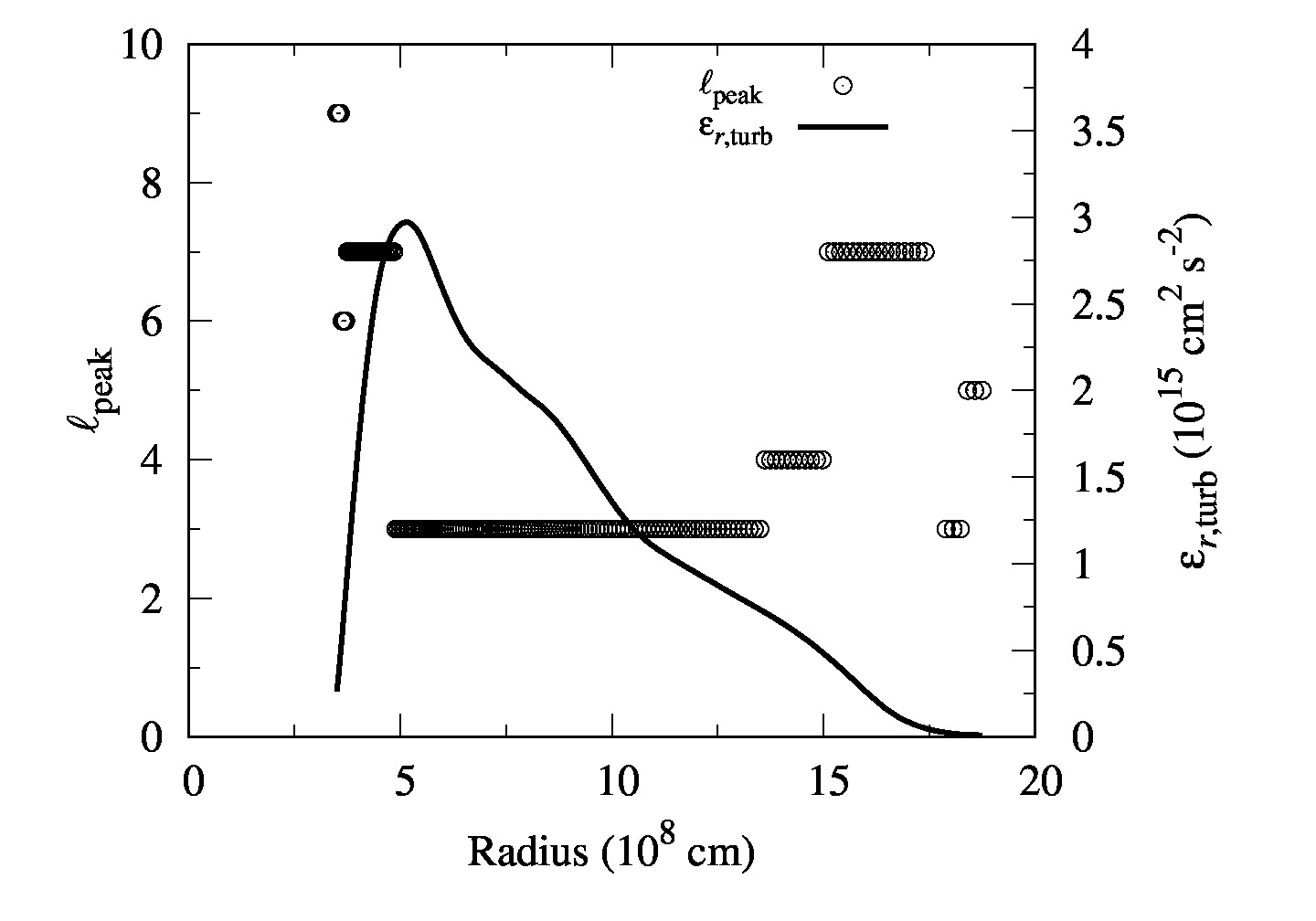}
    \includegraphics[width=0.9 \columnwidth]{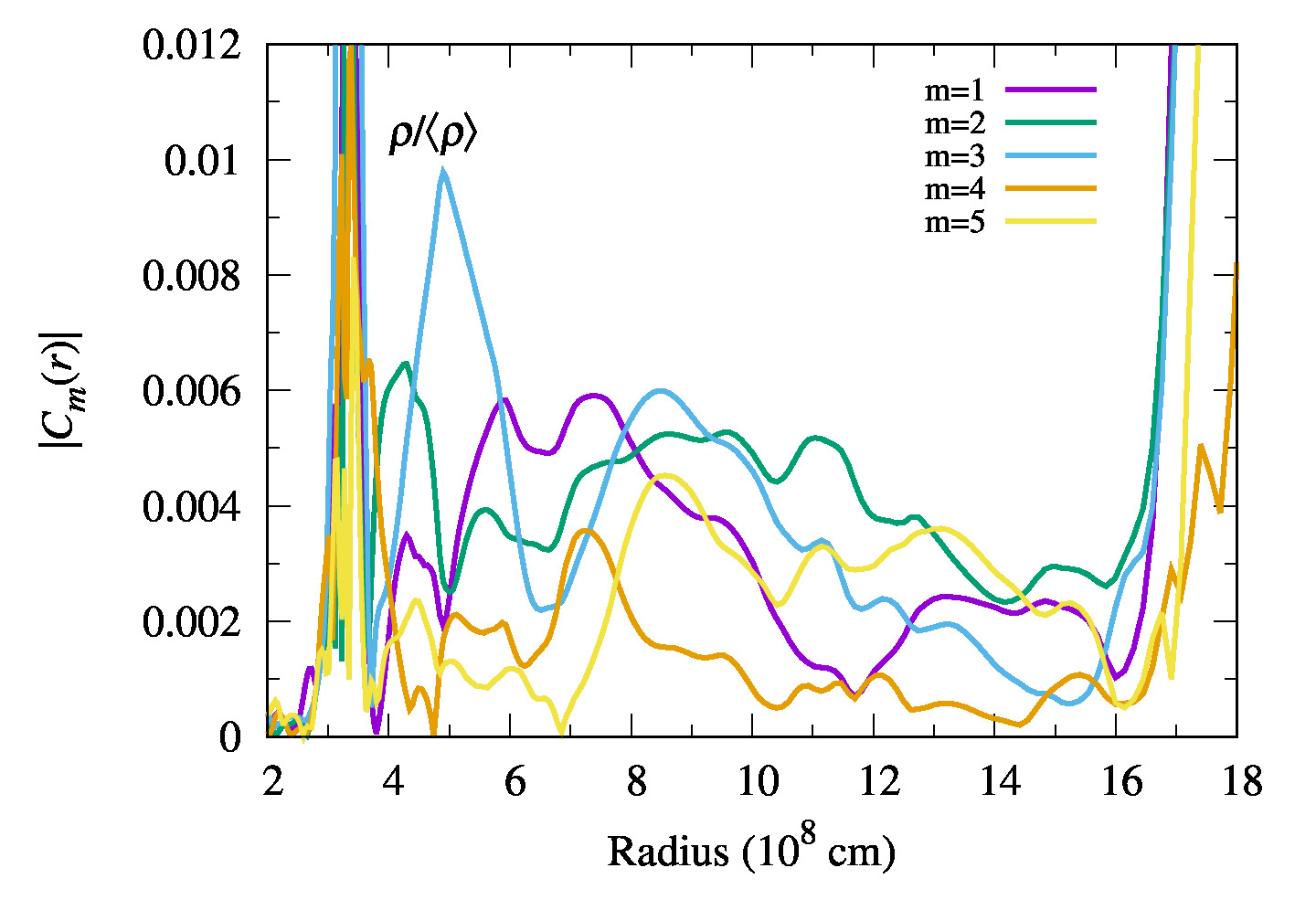}
\caption{The radial profiles of power spectra for the turbulent motion at the last time step.
Top panel: The peak mode $\ell_{\rm peak}(r)$ (circles) and the radial turbulent kinetic energy (curves).
Bottom panel: $|c_m(r)|$ of the density fluctuations.}
\label{fig:power_spectra}
\end{figure}

We calculate the power spectrum of the radial turbulent velocity $c_\ell(r)$ and the specific radial turbulent kinetic energy $\varepsilon_{r,{\rm turb}}(r)$ at the last time step using equations 5 and 6 in \citet{Yoshida21}, respectively.
We show the radial profile of the peak mode $\ell_{\rm peak}(r)$ indicating the maximum value of $c_{\ell}^2(r)$ in the power spectrum.
The top panel of Fig. \ref{fig:power_spectra} shows the radial profiles of the peak mode and the specific radial turbulent energy.
The maximum value of the angle-averaged radial turbulent velocity is $7.7 \times 10^{7}$ cm s$^{-1}$ at $r = 5.2 \times 10^{8}$ cm.
The averaged values of the turbulent velocity and the rotating velocity in the Si/O-rich layer are $5.5 \times 10^7$ and $2.0 \times 10^7$ cm s$^{-1}$, respectively.
The average turbulent velocity is faster than the average rotating velocity.
On the other hand, we expect that the rotating velocity of this star is still about 10 times as fast as the rotating velocity in the Si layer of a typical rotating massive star \citep[e.g.,][]{Heger05}.

We see the dominant mode of $\ell = 3$ in most of the range of the Si/O layer.
This means that large scale turbulent motion is dominant in the Si/O layer.
We evaluate the radially averaged peak mode in the Si/O layer using equation 7 in \citet{Yoshida21}.
The obtained value of the averaged peak mode is $\langle \ell_{\rm peak} \rangle = 3.64$.
This value is similar to the non-rotating massive stars of models 22L, 25M, and 27L$_{\rm A}$ that have been investigated in the previous studies.
Thus, when oxygen-shell burning occurs at the bottom of the Si/O layer, the burning causes large scale turbulence with low modes as $\ell \sim 3$.

Next, in order to clarify what order modes dominate in these turbulent structures, we investigate the power spectrum of order $m$ of the density structure in the {\it x--y} plane.
We calculate the power spectrum of the density structure in the {\it x--y} plane as
\begin{equation}
    C_m(r) = \int_{0}^{2\pi} \rho(r,\phi) \exp(-i m\phi) {\rm d}\phi / \langle \rho \rangle(r) .
\end{equation}

The bottom panel of Fig. \ref{fig:power_spectra} shows the power spectra of order $m$, $|c_m(r)|$ of the density structure in the {\it x--y} plane.
We see a peak of $m = 3$ around $r \sim 5 \times 10^{8}$ cm.
This mode is the same number as the dominant degree mode $\ell$ determined by the turbulent motion.
Since the location of the $m = 3$ is almost same as that of the highest radial turbulent velocity, this mode is considered to be due to the turbulence caused by the oxygen-shell burning.

We also see a broad peak of the $m = 2$ mode in the range $r$ = (7--13)$\times 10^{8}$ cm of the Si/O layer.
We see high density structures in the southeast and north-west direction and low ones in the south and north direction in the right-hand panel of Fig. \ref{fig:2d_e3_rho}.
This spiral arm structure would correspond to the $m = 2$ mode shown in Fig. \ref{fig:power_spectra}.
There is another dominant mode of $m = 1$ in the range $r = $(6--8)$\times 10^{8}$ cm.
These dominant low modes for $m$ are not seen in the power spectrum of the $\ell$ mode.
Thus, it is likely that the low $m$ feature is an effect of rotation in the turbulent motion.
This structure would be formed through differential rotation, i.e., the outward decrease in the angular velocity in this layer, as shown below.
We are not sure whether the above rotating feature is seen in typical rotating massive stars.
To clarify this, we need to investigate 3D simulations for various rotating stars.

\subsection{Profile of specific angular momentum}

\begin{figure}
    \centering
    \includegraphics[width=7.2cm]{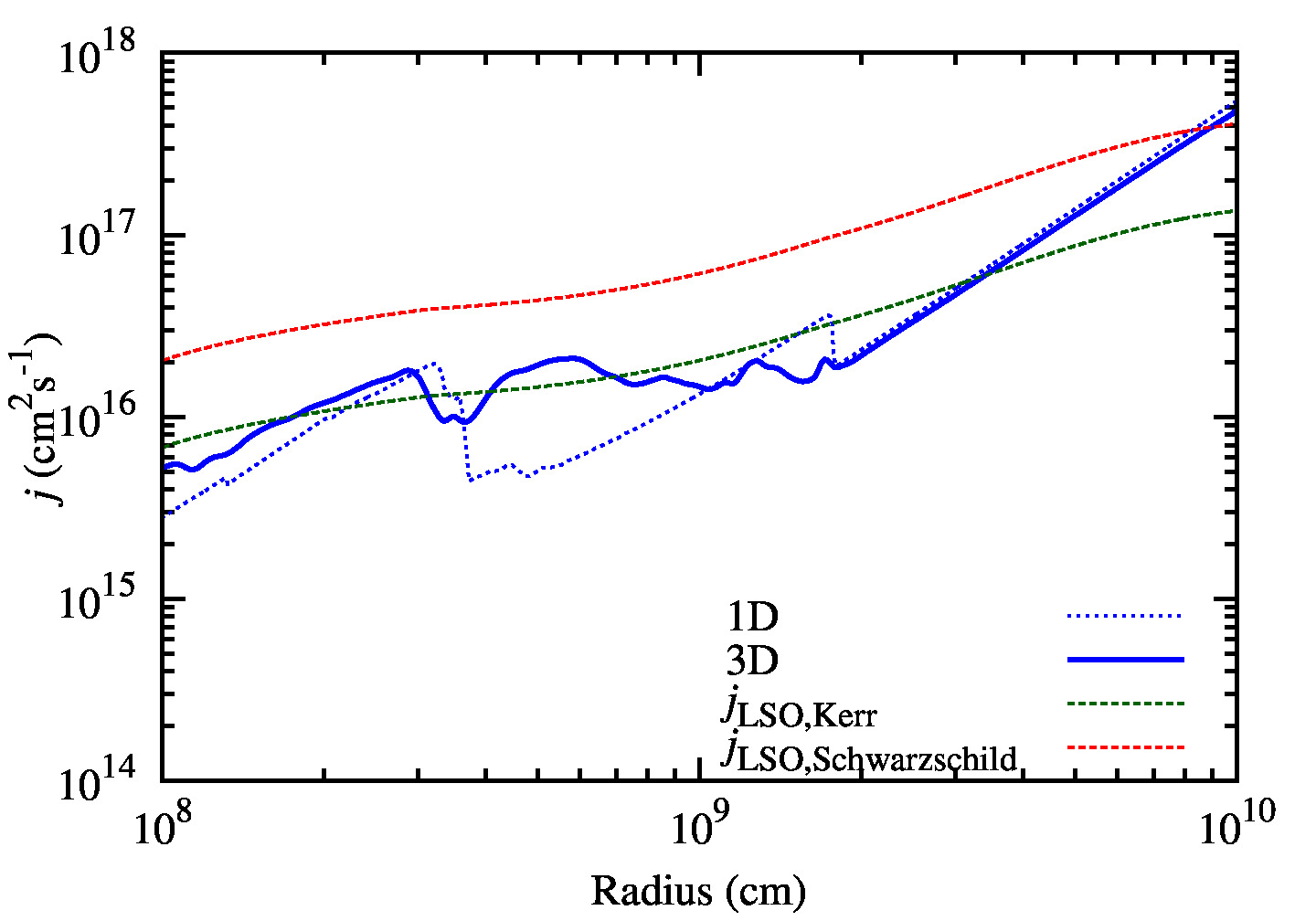}
\caption{Radial profiles of angle-averaged specific angular momentum at the last time step.
Blue dotted and solid curves indicate 1D \citep{Aguilera-Dena20} and 3D results, respectively.
Green and red dashed curves are the angular momentum required to form a stable disc at the last stable orbit (LSO) of a Kerr black hole and a Schwarzschild black hole.
}
\label{fig:omg_j}
\end{figure}

The angular momentum in the Si/O layer is transported by the convective flow.
Here, we investigate the time evolution of radial profiles of the specific angular momentum.
Fig. \ref{fig:omg_j} shows the radial profiles of the angle-averaged specific angular momentum at the last time step.
In the Si/O layer, the specific angular momentum is initially $\propto r^2$ since  the
angular velocity is assumed to be constant in 1D models.
After the 3D simulation, the specific angular momentum becomes close to uniform.
This trend is different from the angular momentum transport in the convection layer of 1D rotating star models.

In 1D evolution calculations of rotating stars, the angular velocity profile in convective layers is treated as rigid rotation \citep[][]{Endal78,Pinsonneault89} or tends to move to a rigid rotation profile \citep[e.g.,][]{Heger00,Meynet00}.
In convective layers, the high viscosity associated by turbulence may lead to rigid rotation \citep{Endal76}.
However, the angular momentum transport in the convective layers has not been totally clarified and other possibilities such as uniform specific angular momentum is also discussed \citep[e.g.,][]{Potter12}.
\citet{Heger00} discussed that rigid rotation seems to be justified, at least if the rotational period is long in compared to the convective time-scale and if convective blobs can be assumed to scatter elastically \citep{Kumar95}.

In our simulation, strong turbulence induced by oxygen-shell burning grows violently from $\sim$70 s and the turbulence has not fully developed before the core collapse.
In regions with small convective perturbations in the linear regime, the angular momentum distribution tends to become uniform
\citep{Kumar95}.
Turbulence occurring for a short period may cause such an angular momentum transport.
In order to assess the detailed angular momentum transport by convection in the Si/O layer, longer-time simulation of the convective motion may be needed.
The angular velocity distribution in convective layers may be uniform for most of the evolution period.
When a fast-rotating massive star finally contracts to collapse, however, violent shell burning may develop turbulence within a short time-scale.
This final turbulence may change the angular velocity distribution in the convective layer from a rigid rotation.
This would be possible for the Si and/or O-rich layers because violent shell burning by contraction may occur.
If this kind of angular momentum transport occurs in earlier stages, outward transport of the angular momentum will be suppressed and the stars might keep larger angular momentum.
Recently, a 3D magnetohydrodynamic simulation of oxygen-shell burning has been investigated \citep{Varma21}.
Although they found that magnetic fields do not appreciably alter the convective flow, the fields might affect the angular momentum transport in convection layers.
We would like to investigate the effects of a magnetic field on the angular momentum transport in the future.

Finally, we compare the specific angular momentum to that of the LSO of a Kerr black hole and a Schwarzschild black hole \citep[e.g.,][]{Woosley12}.
The specific angular momentum of the star exceeds the LSO of Kerr and Schwarzschild black holes at $M_r$ = 11.7 M$_\odot$ ($r = 3.4 \times 10^{9}$ cm), and $M_r$ = 25.4 M$_\odot$ ($r = 9.0 \times 10^{9}$ cm).
Thus, if this star collapses to a black hole, an accretion disc larger than $\sim$3.3 M$_\odot$ would form and the disc material may be ejected as a GRB.
The region that forms an accretion disc is in the O/Ne layer
and is solely determined by the 1D evolution.

\section{Summary}

We have performed a 3D hydrodynamics simulation of a fast-rotating massive star during the last $\sim$90 s before core collapse.
Our stellar model has an initial mass of 38 M$_\odot$, a metallicity of $\sim$1/50 Z$_\odot$, and an initial rotating velocity of 600 km s$^{-1}$.
Turbulent flows in the Si/O layer with $r \sim 5$--17 $\times 10^{8}$ cm are developed with a turbulent Mach number of $\sim$0.135 at maximum and a low-degree mode  ($\ell \sim 3$) dominance.

In the Si/O layer, silicon-rich material produced through oxygen-shell burning moves outward and forms a spiral arm structure in the equatorial plane.
Such a spiral arm structure is also seen for the profiles of the rotational kinetic energy density and the density.
The power spectrum of order $m$ for the density structure in the {\it x--y} plane indicates that low modes such as $m \le 3$ are dominant in most of the Si/O layer.
For the first time, rotation-induced spiral flows are identified in the context of 3D pre-SN models.
The turbulent flows in the Si/O layer transforms the angular momentum distribution from rigid rotation to roughly constant specific angular momentum.
This feature is different from the one seen in the evolution of 1D rotating massive star models.
This angular momentum transport is most likely to occur owing to violent turbulent mixing induced by the final shell burning episode in the contracting star.
If such an angular momentum transfer occurs from earlier stages, fast-rotating stars may keep more angular momentum and become more favorable to form SL SNe and long GRBs.

Our result, while preliminary, suggests a significant impact of fast rotation on the explosion or implosion of the star.
The progenitor asphericities fall in the regime where they may be able to affect the supernova shock.
The asphericities of the $m\sim3$ spiral mode may enhance the neutrino driven convection or spiral standing accretion shock instability \citep[similar to][]{bernhard16_prog,KazTakahashi16}.
That may foster the shock revival or delay the formation of the black hole.

\section*{Acknowledgements}

We thank the referee, Raphael Hirschi, for very careful reading this manuscript and providing valuable comments.
This study was supported in part by the Grant-in-Aid for the Scientific Research of Japan Society for the Promotion of Science (JSPS) KAKENHI grant numbers (JP17H05206, JP17K14306, JP17H01130, JP17H06364, JP18H01212, JP20H05249), by the Central Research Institute of Explosive Stellar Phenomena (REISEP) of Fukuoka University and the associated project (207002).
DRAD was supported by the Stavros Niarchos Foundation (SNF) and the Hellenic Foundation for Research and Innovation (H.F.R.I.) under the 2nd Call of ``Science and Society" Action Always strive for excellence -- ``Theodoros Papazoglou" (Project Number: 01431).
The numerical simulation was done using XC50 at the Center for Computational Astrophysics at National Astronomical Observatory of Japan.

\section*{Data availability}

Result data will be shared on reasonable request to the corresponding author.




\bibliographystyle{mnras}
\bibliography{ms.bbl} 








\bsp	
\label{lastpage}
\end{document}